\documentclass[journal=ancac3,manuscript=article,layout=twocolumn]{achemso}

\setkeys{acs}{articletitle = true} 

\usepackage{natbib}
\usepackage{chemformula} 
\usepackage[T1]{fontenc} 
\usepackage{float} %
\usepackage{seqsplit} 
\usepackage{graphicx}
\usepackage{booktabs} 
\usepackage{multirow}
\usepackage{color, colortbl}
\usepackage[para,flushleft]{threeparttable}
\usepackage{ulem}
\usepackage[super]{nth}
\usepackage{amsmath}
\usepackage[font=scriptsize,labelfont=bf]{caption}
\usepackage[switch]{lineno}
\usepackage{breqn}
\usepackage{makecell} 
\DeclareUnicodeCharacter{2009}{\,} 

\author{Hyeong Jin Kim}
\affiliation{Ames National Laboratory, and Department of Chemical and Biological Engineering, Iowa State University, Ames, Iowa 50011, United States}
\author{Binay P. Nayak}
\affiliation{Ames National Laboratory, and Department of Chemical and Biological Engineering, Iowa State University, Ames, Iowa 50011, United States}
\author{Honghu Zhang}
\affiliation{Center for Functional Nanomaterials and NSLS-II, Brookhaven National Laboratory, Upton, New York 11973, United States}
\author{Benjamin M. Ocko}
\affiliation{NSLS-II, Brookhaven National Laboratory, Upton, New York 11973, United States}
\author{Alex Travesset}
\affiliation{Ames National Laboratory, and Department of Physics and Astronomy, Iowa State University, Ames, Iowa 50011, United States}
\author{David Vaknin}
\affiliation{Ames National Laboratory, and Department of Physics and Astronomy, Iowa State University, Ames, Iowa 50011, United States}
\author{Surya K. Mallapragada}
\email{suryakm@iastate.edu}
\affiliation{Ames National Laboratory, and Department of Chemical and Biological Engineering, Iowa State University, Ames, Iowa 50011, United States}
\author{Wenjie Wang}
\affiliation{Division of Materials Sciences and Engineering, Ames National Laboratory, U.S. DOE, Ames, Iowa 50011, United States}
\email{wenjiew@ameslab.gov}

\title {Two-dimensional assembly of nanoparticles grafted with charged-end-group polymers}

\begin{document}
\begin{abstract}

\noindent\textbf{Hypothesis} 
Introducing charged terminal groups to polymers that graft nanoparticles enables Coulombic control over their assembly by tuning pH and salinity of aqueous suspensions.  

\noindent\textbf{Experiments} 
Gold nanoparticles (AuNPs) are grafted with poly (ethylene glycol) (PEG) terminated with \ch{CH3} (charge neutral), \ch{COOH} (negatively charged) or \ch{NH2} group (positively charged), and characterized with dynamic light scattering, $\zeta$-potential, and thermal gravimetric analysis. Liquid surface X-ray reflectivity (XR) and grazing incidence small-angle X-ray scattering (GISAXS) are used to determine the density profile and in-plane structure of the AuNP assembly across and on the aqueous surface.
\noindent\textbf{Findings} 
Assembly of PEG-AuNPs at the liquid/vapor interface is tunable by adjusting pH or salinity for COOH but less for \ch{NH2} terminals. The distinct assembly behaviors are attributed to the overall charge of PEG-AuNPs as well as PEG conformation. COOH-PEG corona is the most compact and leads to smaller superlattice constants. The net charge per particle depends not only on the PEG terminal groups, but also on the cation  sequestration of PEG and the intrinsic negative charge of AuNP surface. The closeness to overall charge neutrality, and the hydrogen bonding in play, brought by \ch{NH2}-PEG,  enables \ch{NH2}-PEG-AuNPs assembly readily. 

\end{abstract}

\section{Graphical Abstract}
\begin{figure} [!htb]
    \centering
    \includegraphics[width=0.9\linewidth] {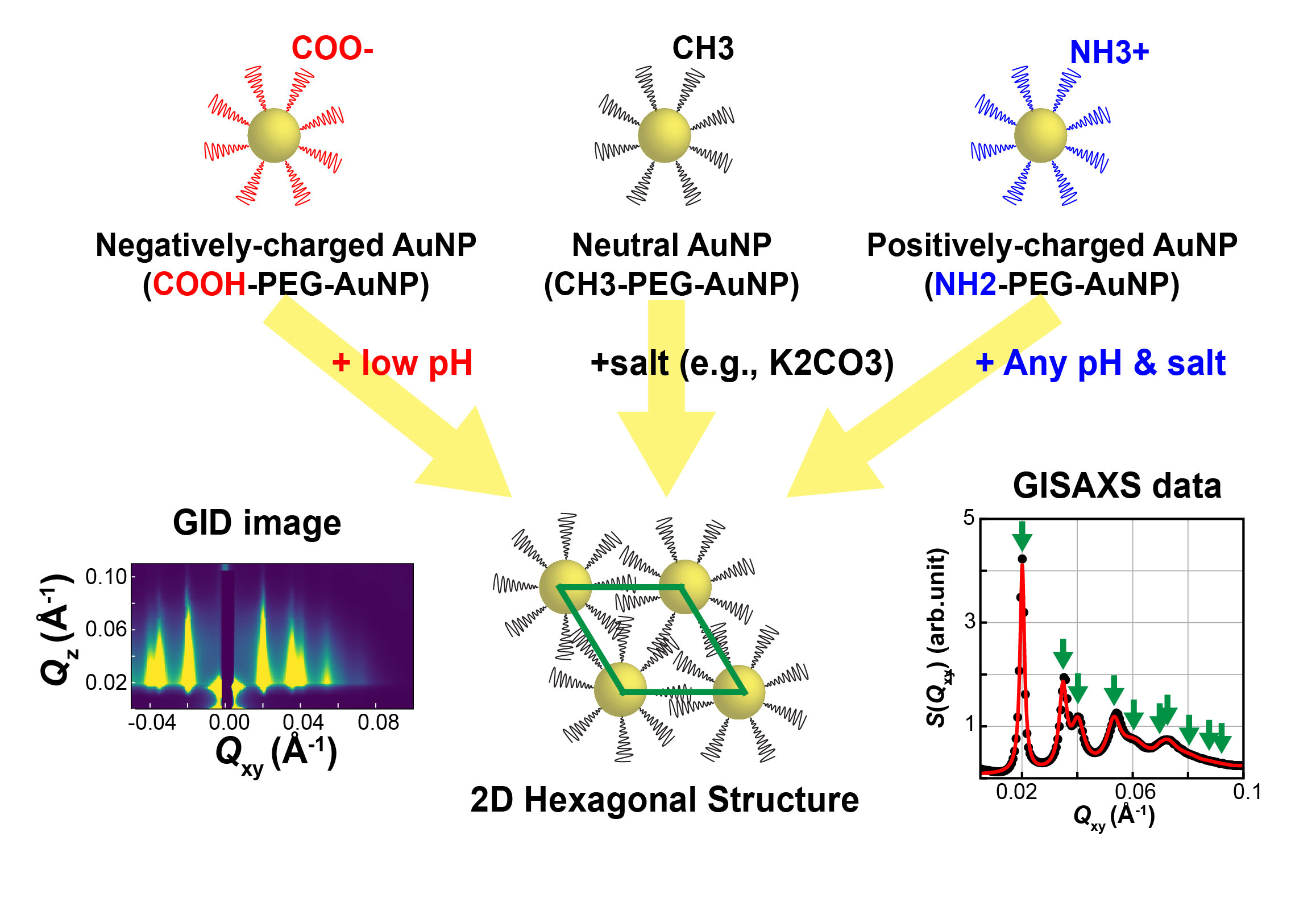}
\label{toc}
\end{figure}

\section{Introduction}
Inorganic nanoparticles (NPs) have attracted great attention due to their unique emergent properties, which are distinct from those based on organic molecules.\cite{andres1996coulomb, chen1998gold, perenboom1981electronic, link1999size}  For example, localized plasmon resonance, superparamagnetism, and photon upconversion can only be achieved with inorganic nanoparticles. \cite{lu2007magnetic, haase2011upconverting, zhan2018plasmon} Self-assembly of such NPs offers a route to fabricating novel devices based on metamaterials frameworks. \cite{daniel2004gold, fan2010self, talapin2010prospects} A host of properties of such NP assemblies are not only dependent on the composition of the NPs, their sizes and shapes, but also on their collective responses to external stimuli, which can be tuned by their mutual arrangements and  interactions.\cite{Ghosh2007ChemRev}  Soft ligands are often used to functionalize inorganic NPs to induce interactions that can lead to assembly, where  the dominant ones are the van der Waals, Coulombic, and the steric interactions.\cite{Bishop2009small,Si2018Adv, grzelczak2019stimuli}  Over the last two decades, the DNA-mediated self-assembly of colloidal superlattices has matured to the extent that a great number of complex  assembly structures have been designed and built, taking advantage, by design, interparticle  base-pairing  hybridization.\cite{nykypanchuk2008dna, park2020dna, laramy2020crystal, macfarlane2020nanoparticle} The use of synthetic polymers as ligands, in place of DNA to mediate the interparticle interactions, is viewed as a more economic, robust and reliable approach to upscale NP-assembly. To accomplish this, water-soluble polymers, including poly(ethylene glycol) (PEG), poly(acrylic acid) (PAA), and poly(N-isopropylacrylamide) (PNIPAM) have been used as NP surface ligands to mediate both two- and three-dimensional superlattice formation of gold nanoparticles (AuNPs) at the liquid/vapor interface and in aqueous bulk solutions under variable mild solvent conditions (e.g., pH, salinity, and temperature). \cite{zhang2017macroscopic, zhang2017ion,zhang2017assembling, zhang2017interfacial, kim2019salt, kim2020temperature, kim2020interfacial, minier2021poly, kim2021nanoparticle, kim2021effects, kim2021effect, kim2022binary, wang2019PRM} Nevertheless, most  NP superlattices are built with charge-neutral ligand functionalized NPs, where the assembly process is mainly driven by the balance between van der Waals attractive forces and steric conformations among polymeric ligands. In crude analogy, using  single stranded DNA to functionalize metal NPs, where the terminal groups are complementary base-paired, has been shown to produce a plethora of binary superstructures.\cite{macfarlane2020nanoparticle}   To introduce similar complimentary interaction,  we graft AuNPs with polymeric ligands that have charged terminals to create "super-ions", and expect  their assembly behaviors vary in response to changing suspensions conditions (pH or salinity). Our long term goal is to discover binary systems where Coulombic interactions among such super-ions can  be tuned to form colloidal clusters, assemblies, and superstructures  analogous to  ionic crystals.\cite{Kalsin2006JACS, Kastilani2018Langmuir, Bridonneau2020JPCB,Curk2021PRL, Pothukuchi2021Lang}

In this study, we explore the assembly of AuNPs that are functionalized with thiolated PEG  terminated with either carboxylic (\ch{COOH}) or amine (\ch{NH2}) groups.  These end-groups render net surface charge to the functionalized NPs and we exploit them as another means to control  PEG-AuNP assembly and crystallization.  In the past, other controls such as  pH, salinity, temperature, and poly-electrolyte co-solutes have been  used to effectively mediate ordered assembly of charge neutral PEG-AuNPs.\cite{zhang2017macroscopic, zhang2017ion,zhang2017assembling, zhang2017interfacial, kim2019salt, kim2020temperature, kim2020interfacial,  kim2021nanoparticle, kim2021effects, kim2021effect, kim2022binary} Here, we compare the self-assembly on the aqueous surface of suspensions of like-charged PEG-AuNPs with a positive, negative, or neutral charge. We note that  the mechanisms of colloidal assembly of like-charged colloidal particles,  based on the Deryaguin–Landau–Verwey–Overbeek theory (DLVO), have been investigated extensively and are still  debated.\cite{Grier1998Nat, Bowen1998Nat, Wu2018NatCom, Santos2019PRL, Rosenberg2020SM} 

In the present study, PEG-AuNPs accumulate on the aqueous surface and form a two-dimensional (2D) assembly.\cite{zhang2017macroscopic} We utilize synchrotron-based liquid surface X-ray diffraction methods, including X-ray reflectivity (XR) and grazing-incidence small angle X-ray scattering (GISAXS), to determine the 2D films at the vapor/liquid interface. This study adds to previous assembly studies using neutral PEG as a surface modifier of the AuNPs.  

\section{Experimental Section}
\subsection{Preparation of Materials}
\begin{figure} [!htb]
    \centering
    \includegraphics[width=0.9\linewidth] {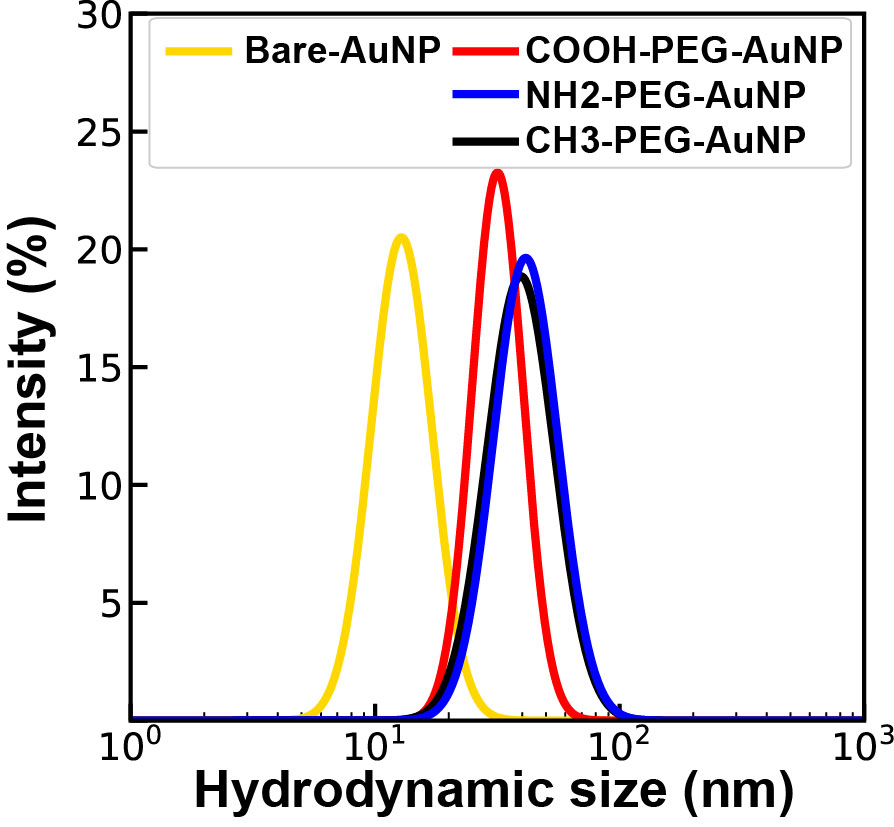}
    \caption{DLS intensity percentage versus hydrodynamic size distribution for the suspensions of bare surface and polymer grafted AuNPs in the absence of salts.}
\label{fig:DLS}
\end{figure}

Citrate-stabilized gold nanoparticles (AuNPs) of $\sim 10$ nm diameter were purchased from Ted Pella. Thiolated poly(ethylene glycol) (HS-PEG; molecular weight of $\sim 5$ kDa) terminated with carboxyl (\ch{COOH}-), methyl (\ch{CH3}-), or amine (\ch{NH2}-) group were obtained from Creative PEGWorks. AuNPs were functionalized with HS-PEG-COOH, HS-PEG-\ch{CH3}, or HS-PEG-\ch{NH2} by ligand-exchange protocol.\cite{kim2022binary}
Briefly, thiolated PEG was dissolved in aqueous solution of sodium chloride (NaCl; 1M). Then, it was incubated with AuNP suspensions for 2 to 3 days. PEG grafted AuNPs (PEG-AuNPs) were purified by centrifugation  three times. In this study,  we use PEG-AuNP to denote PEG grafted AuNPs in general, and \textit{x}-PEG-AuNP to specify the PEG ligands of AuNPs terminated with \textit{x} group (i.e., \ch{CH3}, \ch{COOH}, or \ch{NH2} groups). 
For example, COOH-PEG-AuNPs refers to gold nanoparticles functionalized with carboxyl-terminated PEG.  The  PEG-AuNP concentrations  were determined by Ultraviolet–visible (UV–vis) absorption spectroscopy measurements (calibrated and standardized with vendor provided information, Molecular Devices, SpectraMax M3)  and adjusted to be at approximately 20 nM. Further dynamic light scattering (DLS) characterization of the grafted AuNPs was conducted with a NanoZS90 and its associated software Zetasizer (Malvern, U.K.). The DLS intensity percentage versus particle hydrodynamic size distribution profiles clearly show  an increase in size upon grafting with all three PEGs. (see Table \ref{tbl:dls-zeta} and Figure \ref{fig:DLS}). Compared to  \ch{CH3}-PEG-AuNPs, the  hydrodynamic size of \ch{COOH}-PEG-AuNPs is 18\% smaller  while that of \ch{NH2}-PEG-AuNPs is 5\% larger. Zeta ($\zeta$) potential measurements, also conducted with NanoZS90, were used to quantify the surface charge of PEG-AuNPs as listed in  Table \ref{tbl:dls-zeta}. 
The electrical polarity of COOH-PEG-AuNPs and NH2-PEG-AuNPs, along with enlarged hydrodynamic size, ensures  successful grafting of PEG with distinct terminal groups.

To determine the average grafting density of PEG on each particle surface, thermogravimetric analysis (TGA) of PEG-AuNPs was conducted with Netzsch STA449 and its associated software Proteus. The grafted NPs are degraded in a temperature range between 100 to 600 $^{\circ}$C under argon gas. PEG on AuNP surfaces was found to be thermally degraded between 350 to 450 $^{\circ}$C (shown in Fig. \ref{fig:TGA}).\cite{kim2022binary} 
The grafting density for each sample was calculated using Eq.~\ref{Eq:TGA} and is summarized in table 1 \cite{Benoitgrafting}. 

\ch{K2CO3} stock solutions were prepared
at a sufficiently high concentrations that could be added in small amounts to the grafted nanoparticle suspensions at specified concentrations for the X-ray scattering experiments (see Supporting Information for pH determination).  The HCl, NaOH, and \ch{K2CO3} solutions are treated as ideal solutions and their ionic strength is calculated as  $\frac{1}{2}\sum_{i}c_{i} Z_{i}^2$, where $c_{i}$ and $Z_{i}$ are the concentration and valance of ions of species $i$.

\begin{table*}[!htb]
\caption{\small DLS, $\zeta$ potential and Grafting density values}
\begin{threeparttable}
 \begin{tabular}[htp]{ccc|ccc}
      \toprule \toprule
      \multicolumn{3}{c|}{DLS} & \multicolumn{3}{c}{TGA}\\
      \midrule
      {\scriptsize End group of  } & {\scriptsize   Hydrodynamic}& {\scriptsize $\zeta$ potental } & {\scriptsize Initial weight}  & {\scriptsize Changed } & {\scriptsize   Grafting density}  \\
      {\scriptsize  grafted PEG } & {\scriptsize size (nm)\tnote{1}}& {\scriptsize (mV)} &  {\scriptsize(mg)} & {\scriptsize weight\tnote{3}} &{\scriptsize (chains/nm$^2$)}\\
      \midrule
      {\scriptsize n.a.\tnote{2}} &{\scriptsize 12.7(1)} &{\scriptsize $-38.5\pm2.3$} & - & - & - \\
      {\scriptsize -CH$_{3}$} &{\scriptsize 39.1(2)} &{\scriptsize $-14.5\pm1.8$} & {\scriptsize 1.98} &{\scriptsize 30\%} &{\scriptsize$1.6\pm0.05$}\\
      {\scriptsize -COOH} &{\scriptsize 31.8(3)} &{\scriptsize $-29.4\pm2.1$} & {\scriptsize 2.05} &{\scriptsize 30\%} &{\scriptsize $1.5\pm0.10$}\\
      {\scriptsize -NH$_{2}$} &{\scriptsize 41.0(4)} &{\scriptsize $+13.6\pm2.6$} & {\scriptsize 1.58} &{\scriptsize 32\%} &{\scriptsize $1.7\pm0.15$}\\
      \bottomrule \bottomrule
	\end{tabular}
	\begin{tablenotes}   
	\item [1] {\scriptsize Only the modal size on the distribution profile is reported.}\item [2] {\scriptsize The bare surface AuNPs, stabilized with citrate ligands. 
	
	\item[3] {\scriptsize{The changed weight was calculated from Fig. \ref{fig:TGA} and is relative change with respect to initial weight.}}}
	\end{tablenotes}
\end{threeparttable}
\label{tbl:dls-zeta}
\end{table*}

\subsection{X-ray Experimental Setup}
\textit{In-situ} liquid surface X-ray scattering experiments were carried out at the SMI beamline open platform liquid surfaces (OPLS) end station at the National Synchrotron Light Source II (NSLS-II), Brookhaven National Laboratory,  with incident X-ray energy  9.7 keV. X-ray reflectivity (XR) and grazing-incidence small angle X-ray scattering (GISAXS) were used to investigate the 2D assembled films at the liquid/vapor interface using area detectors. Data correction and reduction were conducted on-site with the beamline routines. Experimental setups and more details can be found elsewhere. \cite{zhang2017macroscopic, kim2022binary}
Below is a brief description of the methodology.
The X-ray reflectivity (XR), $R$, is measured as a function of $Q_z$ that is the vertical component of the scattering vector, $\vec{Q}$, and is normalized to the calculated Fresnel reflectivity, $R_{\rm F}$, for the aqueous solutions. Calculated $R/R_{\rm F}$ profiles to fit the data are obtained and refined with a parameterized multi-slab model that generates the electron density (ED) profile along the surface normal (i.e., depth in $z$-direction), $\rho(z)$,  via Parratt's method that best fits the data.\cite{Nielsen2011}

The grazing incidence small-angle scattering (GISAXS) intensity data is recorded as a two-dimensional function of $Q_{xy}$ and $Q_{z}$, where $Q_{xy}$ is the horizontal component of $\vec{Q}$. A linecut intensity profile of GISAXS, as a function of $Q_{xy}$ is integrated over $Q_z$ range $0.01$--$0.03$ \AA$^{-1}$, denoted as $I(Q_{xy})$. The Bragg-reflection peaks on $I(Q_{xy})$ profiles are associated with a real-space, two-dimensional lattice. The linecut profile $I(Q_{xy})$ divided by the calculated form factor intensity profile by a single nanoparticle,\cite{zhang2017assembling, zhang2017macroscopic, kim2022binary} is denoted as $S(Q_{xy})$, which is proportional to the structure factor that describes the spatial correlation among particles.\cite{Nielsen2011} 

\section{Results and Discussion}
In this section, we compare the surface X-ray scattering results to determine the impact of the charge of the PEG terminal group on aqueous surface NP assembly at room temperature.  By tuning the salinity or pH of the aqueous suspensions and based on previous results, we expect the PEG-AuNPs to populate  the aqueous surface and assemble into ordered lattices.
The XR measurements from the aqueous suspensions provide the electron densities of the enriched surfaces, while the GISAXS measurements provide in-plane NPs arrangement information.\cite{zhang2017macroscopic}
\begin{figure}[!ht]
\centering 
\includegraphics[width=\linewidth]{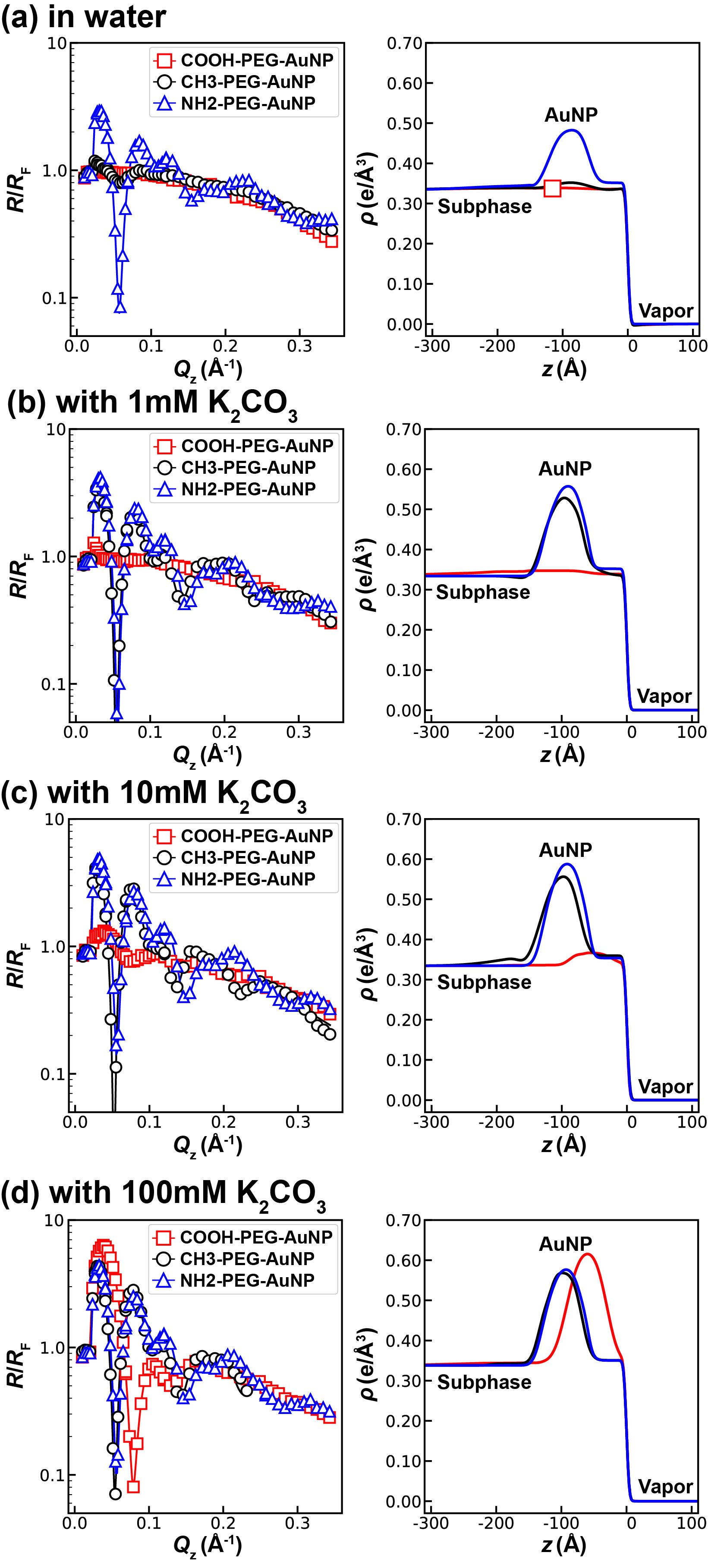}
\caption{ Left: $R/R_{\rm F}$ data \ch{COOH}-, \ch{CH3}-, \ch{NH2}-PEG-AuNPs under various solvent conditions as indicated. Right: Their corresponding electron density (ED) profiles are provided.}
\label{main_ref}
\end{figure}
 
 \begin{figure*}[!hbt]
\centering 
\includegraphics[width=\linewidth]{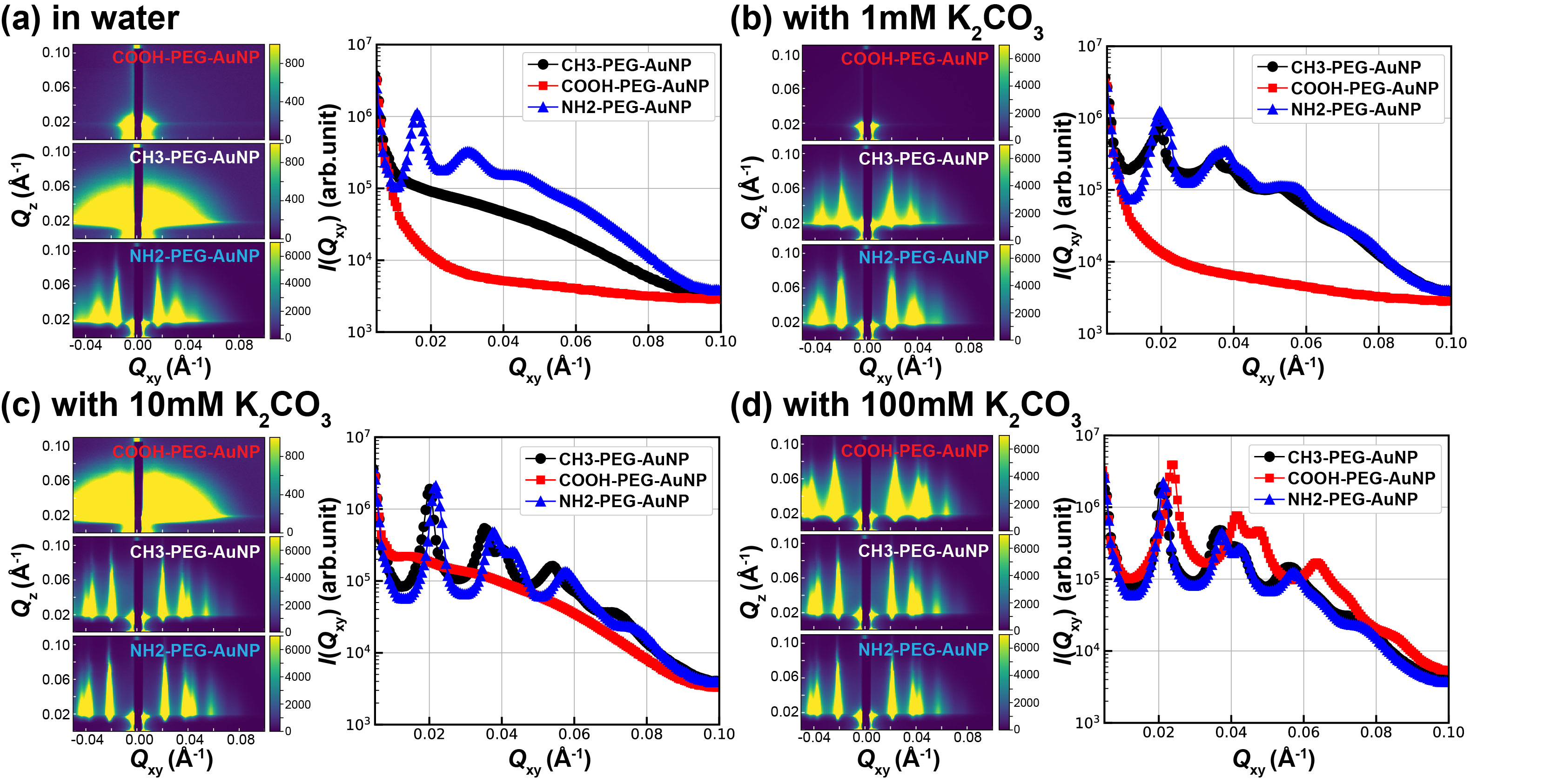}
 \caption{\scriptsize (a-d) Left: GISAXS patterns for \ch{COOH}-, \ch{CH3}-, \ch{NH2}-PEG-AuNPs under various solvent conditions as indicated. Right: Their horizontal linecut intensity profiles integrated over $Q_z$ range $0.01-0.03$ \AA$^{-1}$ are plotted on a logarithmic scale.}
 \label{maink2co3} 
 \end{figure*}
 
\subsection{1. Salt induces 2D assembly.}
 \ch{K2CO3} is among the most effective salts to induce ordered assembly of AuNPs grafted with neutral PEG ligands, \ch{CH3}-PEG, in aqueous suspensions.\cite{zhang2017macroscopic}  This is generally attributed to the salting-out of PEG in response to a poorer solvent enriched with salts. \cite{zhang2017assembling,zhang2017interfacial,zhang2017macroscopic} 
Figure \ref{main_ref} shows normalized  reflectivities, $R/R_{\rm F}$, for PEG-AuNPs in aqueous solutions at various \ch{K2CO3} concentrations. In pure aqueous suspensions without salt, i.e.,  [\ch{K2CO3}]$=$0 mM (Fig. \ref{main_ref}(a)),  the $R/R_{\rm F}$ profiles of \ch{CH3}-PEG-AuNPs and \ch{COOH}-PEG-AuNPs suspensions fall off  monotonously with $Q_z$,  similar to that of a bare and smooth liquid/vapor interface. The lack of $Q_z$-dependent modulation does not support surface NP adsorption. Meanwhile, the $R/R_{\rm F}$ profile for \ch{NH2}-PEG-AuNPs shows a prominent interference pattern. This is  evidence for the spontaneous  formation of a thin film  of \ch{NH2}-PEG-AuNPs at the vapor/liquid interface. This is surprising since the amine group presumably renders a positive charge at the surface of the PEG-corona for each AuNP. It is therefore expected that such charged AuNPs would tend to be fully dispersed (soluble) in aqueous suspensions due to Coulombic inter-particle repulsion. However, these repulsive interactions can be weakened by charge screening of co-ions and the salting-out of PEG. With the addition of 1 mM \ch{K2CO3}, in Fig.\ \ref{main_ref}(b), the $R/R_{\rm F}$ of \ch{CH3}-PEG-AuNPs is very similar to that of  \ch{NH2}-PEG-AuNPs, where the latter remains practically unchanged upon addition of \ch{K2CO3}, i.e, formed film is almost saturated even at 0 mM \ch{K2CO3}.  The negatively-charged \ch{COOH}-PEG-AuNPs exhibits a distinctive interference pattern in $R/R_{\rm F}$ at 10 mM \ch{K2CO3}, with a much stronger effect at 100 mM \ch{K2CO3}, as shown in Fig.\ref{main_ref} (c) and (d).  

The corresponding ED profiles across the interfaces obtained by fitting the $R/R_{\rm F}$  are shown in the right panel of Fig.\ \ref{main_ref}. The bell-shaped peak with high ED represents a uniform layer formed by the electron-rich AuNPs, and the ED excess on both sides of the AuNPs film is attributed to PEG. We note that it is difficult to distinguish the PEG protruding into the aqueous suspension as its ED is close to that of water\cite{zhang2017interfacial}.  A summary of inspections of ED profiles is as follows: 1. The threshold of bulk [\ch{K2CO3}], above which the surface enrichment of PEG-AuNPs are observed, is ranked as: \ch{NH2}-PEG-AuNPs < \ch{CH3}-PEG-AuNPs < \ch{COOH}-PEG-AuNPs. The   \ch{NH2}-PEG-AuNPs spontaneously populate and saturate the aqueous surface without any salt as their surface density is  practically insensitive to the further addition of \ch{K2CO3}. The \ch{CH3}-PEG-AuNPs populate the surface sporadically at [\ch{K2CO3}]=0 mM and vastly after the addition of 1mM \ch{K2CO3}. The \ch{COOH}-PEG-AuNPs require a larger amount of \ch{K2CO3} (10-100 mM) to fully populate the surface. 2. Once the surfaces are saturated, the  ED profiles for \ch{CH3}-PEG-AuNPs and \ch{NH2}-PEG-AuNPs are almost indistinguishable, and both feature a flat ED transition zone between vapor phase ($\rho =0$) and the bell-shaped high-ED zone that is mostly of Au cores.\cite{wang2019PRM} This flat ED zone is attributed to the compressed PEG ligands that are in the immediate vicinity of the vapor phase.\cite{kim2021effect} In contrast,
the lack of such compressed PEG sub-layer for \ch{COOH}-PEG-AuNPs films is reminiscent of previously studied PNIPAM-AuNPs films where the PNIPAM ligands collapse on the AuNP surface upon salt addition and assembly.\cite{minier2021poly}  The scenario that the PEG chains partially  collapse is manifested by the DLS results where the modal hydrodynamic size and size distribution are smaller and narrower for \ch{COOH}-PEG-AuNPs than the other counterpart terminal groups.

\begin{figure}[!ht]
\centering 
\includegraphics[width=\linewidth]{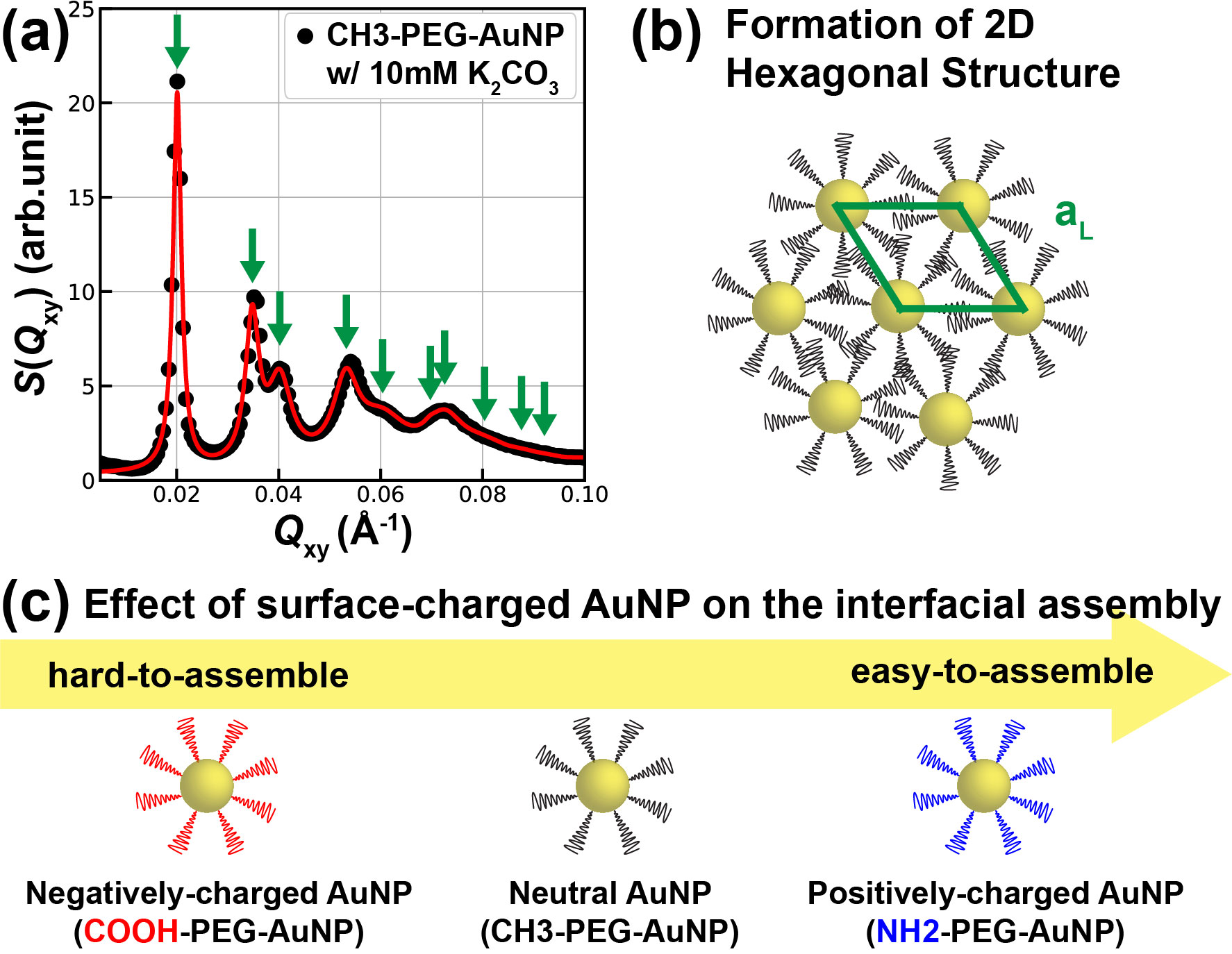}
\caption{\scriptsize (a)   $S(Q_{xy})$ profile proportional to the in-plane structure factor for methyl-terminated PEG-AuNPs at 10 mM \ch{K2CO3} (black circles). Red solid curve is the best-fit using a 2D hexagonal lattice. Green arrows indicate the $Q_{xy}$ positions of Bragg diffraction peaks for 2D hexagonal lattice. (b) Illustration for 2D hexagonal lattice of PEG-AuNPs with a lattice constant, $a_L$. (c) Schematic representation of effects of surface-charged AuNPs on the interfacial assembly.}
\label{main_hex}
\end{figure}

\begin{figure*} 
\centering 
\includegraphics[width=0.8\linewidth]{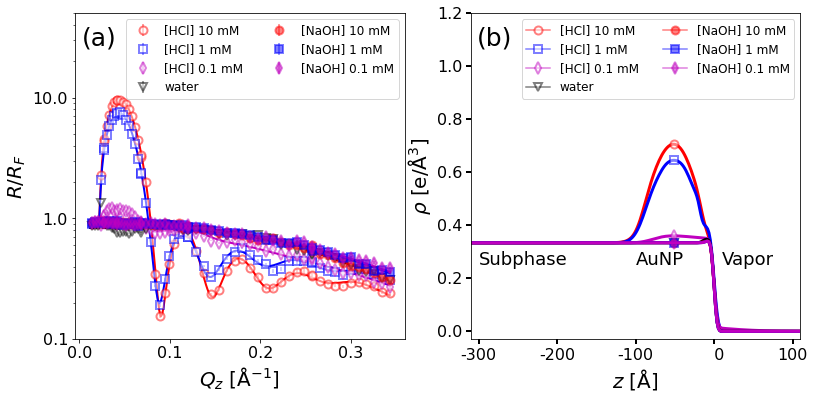}
\caption{(a) $R/R_{\rm F}$ data on logarithmic scale for carboxyl-terminated (COOH-) PEG-AuNPs under various solvent conditions as indicated. (b) Corresponding electron density (ED) profiles that best fit $R/R_{\rm F}$ data in (a).} 
\label{main_ph_ref}
\end{figure*}

To complement the main findings from the XR above, we present and analyze in-plane diffraction results from the same films using GISAXS.  
The GISAXS results are presented using two-dimensional colormaps of the scattering-intensity as a function of $Q_z$ and $Q_{xy}$, shown in Figure \ref{maink2co3}. 
In general, for a bare vapor/liquid interface, the GISAXS map is dominated by the diffuse scattering tail of the reflected X-ray primary beam  and the beam-stop. Typical scattering images are shown in Figure \ref{maink2co3}.  The upper left panel of Fig.\ref{maink2co3}(a) shows the image for \ch{COOH}-PEG-AuNPs without any salt added, indicating a bare surface.  The absence of additional features, except from the direct beam, is consistent with the conclusion from the XR results that the \ch{COOH}-PEG-AuNPs in the absence of salts do not populate the surface.  For \ch{CH3}-PEG-AuNPs without salt the GISAXS map shows a broad  isotropic halo arising from the form-factor of  AuNPs that sparsely populate the surface and lack spatial correlations.  XR is mostly insensitive to such details, but is indirectly manifested by a very weak peak in  $R/R_F$ at low $Q_z$ values that deviates from that of an ideal bare liquid surface (see Fig.\ \ref{main_ref}(a)). 
 For \ch{NH2}-PEG-AuNPs in the absence of salts, the GISAXS pattern  shows a few broad rod-like Bragg-reflections (e.g., Figure \ref{maink2co3}(a) bottom left panel), modulated by the AuNP form-factor mentioned above.  Such rod-like features result from spatially ordered AuNPs confined in 2D films  at the liquid/vapor interface. 

 At [\ch{K2CO3}]= 1 mM, \ch{COOH}-PEG-AuNPs are still dispersed in the bulk of the suspension and do not migrate to the liquid/vapor interface. However, both \ch{NH2}- and \ch{CH3}-PEG-AuNPs at this same salt concentration  show similar but sharper Bragg-reflection rods, consistent with our XR results that show a uniform film formation.  
  At [\ch{K2CO3}]= 10 mM, \ch{COOH}-PEG-AuNPs show the form-factor-like pattern of AuNPs that indicates partial population at the surface, with no correlations among particles (Fig.  \ref{maink2co3}(c)).  This is consistent with the XR data showing excess ED at the surface (Fig. \ref{main_ref}(c)). At this concentration, the \ch{NH2}- and \ch{CH3}-PEG-AuNP  show much sharper and additional Bragg-reflections. 
  At [\ch{K2CO3}]= 100 mM, the images for \ch{COOH}-PEG-AuNPs show a strong diffraction pattern, similar, but shifted to larger $Q_{xy}$ values compared, to the patterns of \ch{NH2}- and \ch{CH3}-PEG-AuNPs systems. The shift of diffraction pattern to larger $Q_{xy}$ suggests shorter inter-particle separations for \ch{COOH}-PEG-AuNPs, in accordance with the thinner or collapsed PEG corona supported by DLS and XR, respectively. 
  
  To quantify the observed patterns, we display, side-by-side, line-cuts form the 2D maps obtained by integrating the intensity over a finite $Q_z$ range (0.01 -- 0.03 {\AA}$^{-1}$). 
  The line-cuts clearly show  AuNP form-factor for \ch{CH3}-PEG-AuNPs at [\ch{K2CO3}]=0 mM  and for \ch{COOH}-PEG-AuNPs at [\ch{K2CO3}]=10 mM.  At these concentrations, the in-plane Bragg reflections of \ch{NH2}-PEG-AuNPs are quite broad, indicating in-plane hexagonal short-range order. As the concentration increases further all three systems show sharper diffraction peaks that can be indexed by a two-dimensional hexagonal lattice where up to 7th order reflection is observed.  For demonstration, Fig. \ \ref{main_hex}(a) shows the in-plane diffraction pattern, $S(Q_{xy})$, from the 2D hexagonal structure of  \ch{CH3}-PEG-AuNPs (at 10 mM \ch{K2CO3}) with arrows marking the calculated Bragg reflection positions for this structure shown in Fig. \ref{main_hex} (b).  Similar structures are also formed by \ch{NH2}-PEG-AuNPs and \ch{COOH}-PEG-AuNPs.  
  Combining GISAXS, XR and DLS results, we conclude that the COOH terminal group induces some kind of partial  polymer collapse, such that the effective radius of the \ch{COOH}-PEG-AuNPs is smaller than the ones grafted with the other terminal groups.  The extracted structural parameters for the XR and the GISAXS are listed in Table \ref{main_tbl}.

\begin{figure*}[!ht]
\centering 
\includegraphics[width=\linewidth]{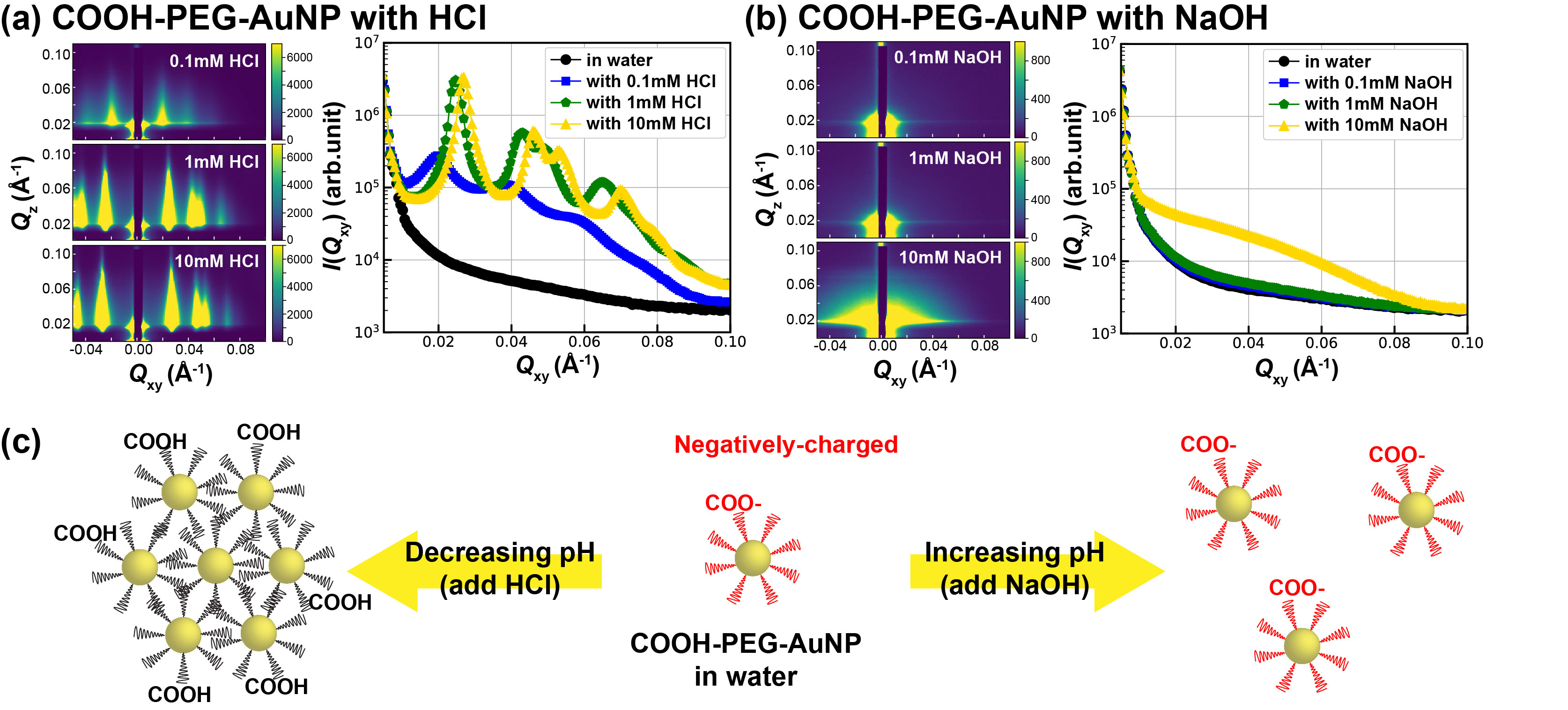}
\caption{ (a-b) Left: GISAXS patterns for carboxyl-terminated PEG-AuNPs under various solvent conditions as indicated. Right: Their horizontal linecut intensity profiles are plotted on a logarithmic scale. (c) Schematic representation of assembling behavior of carboxyl-terminated PEG-AuNPs (negatively-charged) by pH changes.}
\label{main_ph}
\end{figure*}
\subsection{2. pH tunes the threshold for assembly of \ch{COOH}-PEG-AuNPs.}
The effect of pH on the assembly of \ch{CH3}-PEG-AuNPs has been previously investigated. \cite{zhang2017interfacial,Nayak2019Lang} These studies show that tuning pH  with equal [HCl] or [NaOH] has same effect on assembly of \ch{CH3}-PEG-AuNPs.\cite{Nayak2019Lang} 
 

It is expected that lowering the pH of the suspension will result in protonation of the carboxylic terminals, neutralizing the PEG corona shells, and inducing 2D aggregation of charge neutral \ch{COOH}-PEG-AuNPs at the liquid/ vapor interface. In contrast, adding NaOH is expected to promote deprotonation and  more negatively charge the \ch{COOH/COO-}-PEG-AuNPs surface hence favoring solubility. 
Figure  \ref{main_ph_ref} shows $R/R_{\rm F}$ profiles for \ch{COOH}-PEG-AuNPs at different pH levels via the addition of HCl or NaOH. It shows that adding HCl promotes 2D assembly while adding NaOH does not, as expected. We also note that by adding salt or lowering the pH the $R/R_F$  for \ch{COOH}-PEG-AuNPs displays distinct features.  Lowering the pH has a higher first maximum and higher minimum, compared to other two terminal groups with \ch{K2C O3} shown above.  These kind of $R/R_F$ profiles are similar to those observed for AuNPs that are grafted with Poly(N-isopropylacrylamide) (PNIPAM) for which the conclusion drawn was that  the polymer chains partially collapse toward the surface of the core AuNPs\cite{minier2021poly,londono2021salt}, thus allowing for higher surface population of AuNPs.  
On the other hand, $R/R_{\rm F}$ profiles do not exhibit interference ripples for [NaOH] > 0.1 mM, suggesting only marginal surface population of \ch{COOH}-PEG-AuNPs.

\begin{figure}[!ht]
\centering 
\includegraphics[width=\linewidth]{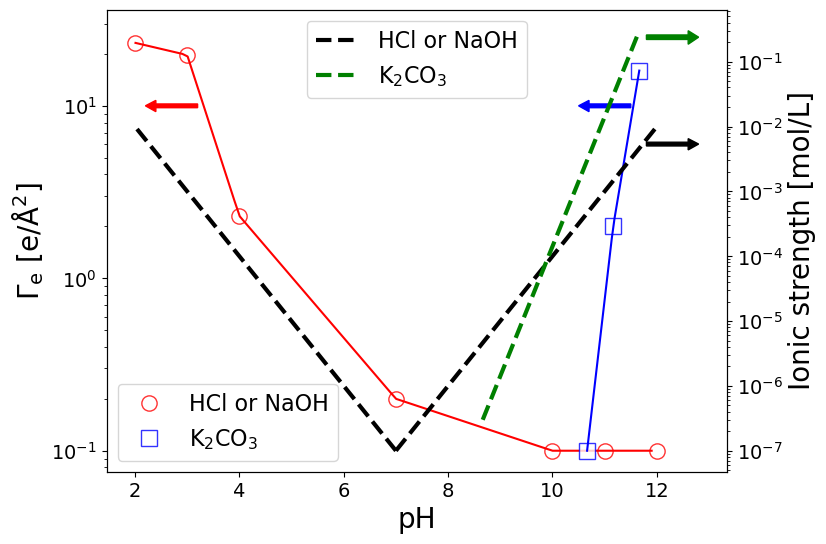}
\caption{ The surface electron excess, $\Gamma_{\rm e}$, as a function of pH tuned by HCl, NaOH, or \ch{K2CO3} for \ch{COOH}-PEG-AuNP suspensions. Dashed lines are calculated ionic strength for HCl, NaOH and \ch{K2CO3} solutions as a function of pH. } 
\label{fig:gamma_COOH}
\end{figure}

\begin{figure*}[!hbt]
 \centering 
 \includegraphics[width=\linewidth]{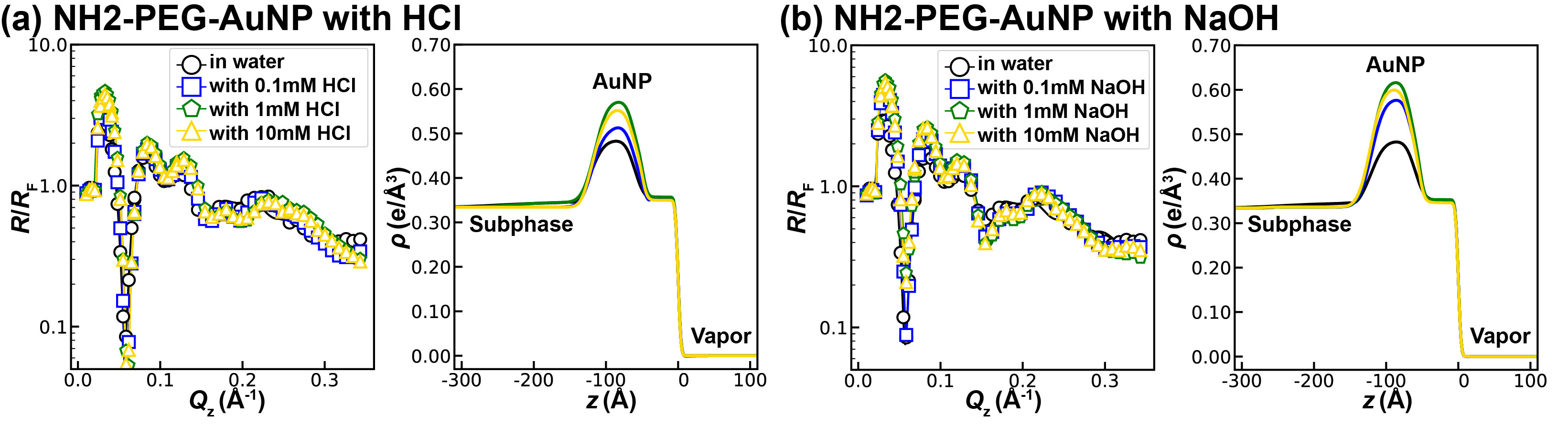}
 \caption{ (a,b) Left: $R/R_{\rm F}$ data for amine-terminated PEG-AuNPs under various solvent conditions as indicated. Right: Their corresponding electron density (ED) profiles are provided.}
 \label{SI_nh2_ph_ref}
\end{figure*}

GISAXS results shown in Figure \ref{main_ph} are consistent with the XR results shown in Fig. \ref{main_ph_ref}. At neutral pH where the suspension has no salt \ch{K2CO3}, HCl, or NaOH, the GISAXS patterns and line-scans mainly originate from the  background, manifested as featureless, falling-off line-cut intensity curves. Upon lowering the pH (i.e., increasing [HCl]),  Bragg rods emerge at pH $4.0$ and become sharper at pH $3.0$ and below. The Bragg rods are at positions expected for a 2D hexagonal lattice of AuNPs at the vapor/ liquid interface. In contrast, at higher pH (i.e., increasing [NaOH]), the GISAXS results resemble that found at neutral pH, suggesting absence of surface enrichment.  However, at the highest pH tested, pH $12.0$ (i.e., [NaOH]=10 mM), the line-cut intensity profile  resembles that of the AuNP form factor, indicating the surface bound AuNPs have accumulated at the air/water interface, albeit without long-range in-plane order.  We speculate that larger pH values may lead to long-range in-plane order.We note that at very large NaOH concentrations, the co-ion, i.e., \ch{Na+}, may bind 
 back to the \ch{COO-} group making the PEG-AuNP surface more close to charge neutral and hence induce particles' migration to the surface. Such a phenomena is  similar to the one observed, but for the oppositely charged amine headgroup monolayer, that at very low pH the co-ion (\ch{Cl-} in this case) starts neutralizing the head group.\cite{Avazbaeva2015} 
 In the Theoretical Considerations section, we explore more probable cause in terms of PEG and ion interactions.

Both XR and GISAXS results suggest that the \ch{COOH}-PEG-AuNPs can be treated as super-ions where their electrostatic interaction strength can be manipulated with pH, as shown in Figure \ref{main_ph}(c). The charged state of the super-ions depends on  deprotonation and protonation on their PEG corona.  At sufficiently high [HCl], the carboxylic terminals are fully protonated and become near charge neutral, and particles assemble in the presence of ions.  Abundance of \ch{Cl-} attributes to poor solvent for PEG, and assembly states are favored as the van der Waals interaction among PEG ligands kicks in. But for pH > 7, the carboxylic terminals are deprotonated and thus charged, leading to repulsive interactions among like-charge super-ions. 
The repulsive interaction trumps the attractive interaction among PEG ligands and counteracts aggregation.
However, the attractive interactions may trump repulsive interaction once ionic contents increase (i.e., \ch{Na+} and \ch{OH-}), as shown in Figure \ref{main_ph}(b).
Indeed, the results with \ch{K2CO3} at similar pH level as NaOH but significantly higher ionic strength also show that \ch{COOH}-PEG-AuNPs assemble at [\ch{K2CO3}]=100 mM, which has orders of magnitude higher ionic strength than 10 mM NaOH (similar pH level). Our comment above about the possibility of of co-ion (e.g., \ch{Na+}) binding at large pH values (i.e., high NaOH concentration) can also be found in Ref.~[\!\!\citenum{Avazbaeva2015}].

In summary, the charged \ch{COOH}-terminals seemingly dictate repulsive and attractive interactions that are tunable by pH and salinity, respectively.  
Figure \ref{fig:gamma_COOH} summarizes the surface electron excess, $\Gamma_{\rm e}$, a measure of surface enrichment of AuNPs, as a function of pH tuned by HCl, NaOH, or \ch{K2CO3} individually. 

\begin{figure*}[!htb]
	\centering 
	\includegraphics[width=\linewidth]{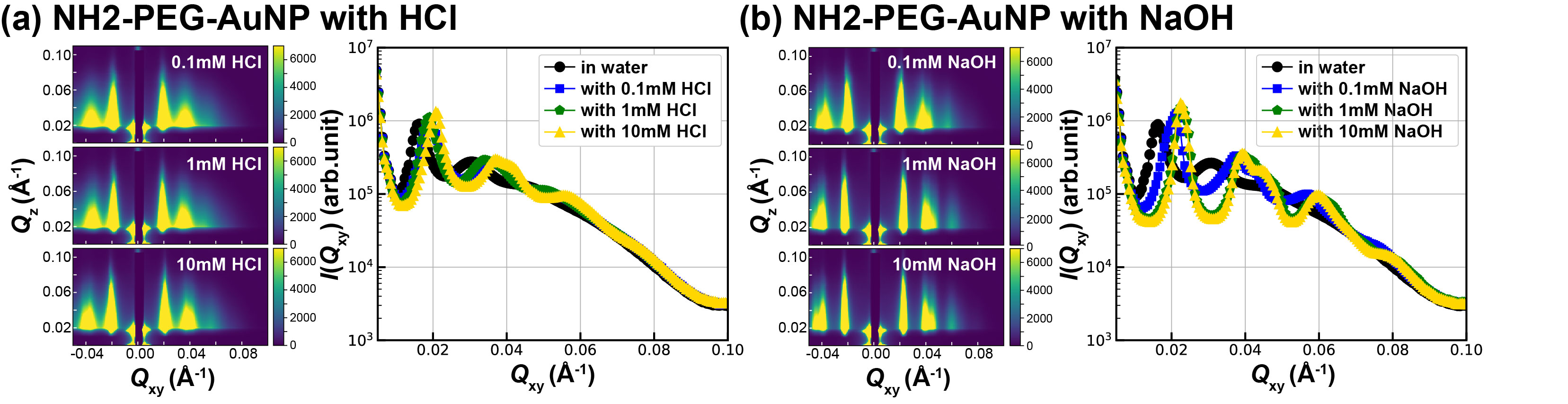}
	\caption{\scriptsize (a-b) Left: GISAXS patterns for amine-terminated PEG-AuNPs under various solvent conditions as indicated. Right: Their horizontal linecut intensity profiles are plotted on a logarithmic scale.
 	}
	\label{SI_nh2_ph} 
\end{figure*}

\subsection{3. pH affects \ch{NH2}-PEG-AuNPs similar to \ch{CH3}-PEG-AuNPs.}
\normalsize


\begin{table*}[!htb]
\centering
\caption{\scriptsize  GISAXS and XR results for carboxyl-terminated (COOH-), methyl-terminated (CH3-), and amine-terminated (NH2-) PEG-AuNPs under various solvent conditions as indicated.}

\begin{threeparttable}
\begin{tabular}{lllcc}
\hline \hline
\makecell{\small End group of \\ grafted PEG} & {\small Solvent condition}& {\small GISAXS results} & {\small $a$\tnote{*} (nm)} & {\small $\Gamma_{\rm e}$\tnote{**}     \ \ (e/{\AA$^2$})} \\
\hline
{\small -CH$_{3}$} & {\small pure water} & {\small Form factor} &{\small n.a.} & {\small < 0.5}  \\
{\small -CH$_{3}$} & {\small 1mM \ch{K2CO3}} & {\small 2D hexagonal structure} &{\small $\sim 37$} & {\small $\sim 12$}    \\
{\small -CH$_{3}$} & {\small 10mM \ch{K2CO3}} & {\small 2D hexagonal structure} &{\small $\sim 36$} & {\small $\sim  17$}    \\
{\small -CH$_{3}$} & {\small 100mM \ch{K2CO3}} & {\small 2D hexagonal structure} &{\small $\sim 34$} & {\small $\sim  16$}    \\
{\small -COOH} & {\small pure water} & {\small Background-like} &{\small n.a.} & {\small < 0.5}  \\
{\small -COOH} & {\small 1mM \ch{K2CO3}} & {\small Background-like} &{\small n.a.} & {\small < 0.5}  \\
{\small -COOH} & {\small 10mM \ch{K2CO3}} & {\small Form factor} &{\small n.a.} & {\small $\sim 2$}    \\
{\small -COOH} & {\small 100mM \ch{K2CO3}}  & {\small 2D hexagonal structure} &{\small $\sim 32$} & {\small $\sim  16$}\\
{\small -NH$_{2}$} & {\small pure water} & {\small 2D hexagonal structure} &{\small $\sim 44$} & {\small $\sim  11$}\\
{\small -NH$_{2}$} & {\small 1mM \ch{K2CO3}} & {\small 2D hexagonal structure} &{\small $\sim 36$} & {\small $\sim  14$} \\
{\small -NH$_{2}$} & {\small 10mM \ch{K2CO3}} & {\small 2D hexagonal structure} &{\small $\sim 33$} & {\small $\sim  17$}\\
{\small -NH$_{2}$} & {\small 100mM \ch{K2CO3}} & {\small 2D hexagonal structure} &{\small $\sim 33$} & {\small $\sim  16$}\\
\hline
{\small -COOH} & {\small 0.1mM \ch{HCl}} & {\small 2D hexagonal structure} &{\small $\sim 38$} & {\small $\sim  2$}\\
{\small -COOH} & {\small 1mM \ch{HCl}} & {\small 2D hexagonal structure} &{\small $\sim 29$} & {\small $\sim  20$}\\
{\small -COOH} & {\small 10mM \ch{HCl}} &{\small 2D hexagonal structure} &{\small $\sim 27$} & {\small $\sim  23$}\\
{\small -COOH} & {\small 0.1mM \ch{NaOH}}& {\small Background-like} &{\small n.a.} & {\small < 0.5}   \\
{\small -COOH} & {\small 1mM \ch{NaOH}} & {\small Background-like} &{\small n.a.} & {\small <0.5}   \\
{\small -COOH} & {\small 10mM \ch{NaOH}} & {\small Form factor} &{\small n.a.} & {\small < 0.5}    \\
{\small -NH$_{2}$} & {\small 0.1mM \ch{HCl}} & {\small 2D hexagonal structure} &{\small $\sim 38$} & {\small $\sim  12$}\\
{\small -NH$_{2}$} & {\small 1mM \ch{HCl}} & {\small 2D hexagonal structure} &{\small $\sim 38$} & {\small $\sim  17$}\\
{\small -NH$_{2}$} & {\small 10mM \ch{HCl}}& {\small 2D hexagonal structure} &{\small $\sim 35$} & {\small $\sim  14$}\\
{\small -NH$_{2}$} & {\small 0.1mM \ch{NaOH}} & {\small 2D hexagonal structure} &{\small $\sim 34$} & {\small $\sim  16$}\\
{\small -NH$_{2}$} & {\small 1mM \ch{NaOH}} & {\small 2D hexagonal structure} &{\small $\sim 32$} & {\small $\sim  19$}\\
{\small -NH$_{2}$} & {\small 10mM \ch{NaOH}} & {\small 2D hexagonal structure} &{\small $\sim 32$} & {\small $\sim  17$}\\
\hline \hline
\end{tabular}
\begin{tablenotes}
\item [*] {\scriptsize For 2D hexagonal structure, lattice constant, $a= 4 \pi /(\sqrt{3} Q_1)$.}
\item [**] {\scriptsize The excess of surface electron density in ED profiles, $\Gamma_{\rm e}=\int[\rho(z)-\rho_{\rm sub}(z)]{\rm d}z$.}
\end{tablenotes}
\end{threeparttable}
\label{main_tbl}
\end{table*}
\ch{NH2}-PEG-AuNPs suspensions can also be expected to respond to pH manipulation, through the addition of HCl or NaOH. Surprisingly, for \ch{NH2}-PEG-AuNPs  lowering or increasing pH levels leads to an increase of the peak in the AuNP ED, as shown in Fig.\ \ref{SI_nh2_ph_ref} (a) and (b). At neutral pH there is spontaneous monolayer formation at the liquid/vapor interface for \ch{NH2}-PEG-AuNPs. There is prominent surface ED enrichment of NPs regardless of pH levels.
This is similar to earlier results for \ch{CH3}-PEG-AuNPs, which show that what is needed to induce assembly are ions (electrolytes) in the suspension.\cite{zhang2017ion}   Fig.\ \ref{SI_nh2_ph} shows GISAXS intensity colormaps and and line-cuts for low pH (with HCl) and at high pH (with NaOH).  The line-cut profiles show that the diffraction pattern shifts to larger $Q_{xy}$ values with either incremental or decremental pH from neutral.  This indicates a decreasing lattice constant of the hexagonal lattice.  Such a decrease in lattice constant is consistent with the increase in surface density determined from XR analysis (See Table \ref{main_tbl}).  Although lowering the pH increases the AuNPs density at the surface, the GISAXS peaks are broader than those for an ideal hexagonal crystal (see Fig.\ \ref{main_hex} where adjacent hexagonal peaks can be resolved). On the other hand, the increase in pH yields higher-quality crystals with better-resolved peaks, indicating the assembly behavior is influenced by amine terminals. Structural parameters extracted from the XR and GISAXS for \ch{NH2}-PEG-AuNPs are listed in Table \ref{main_tbl}.  
 
\section{Theoretical Considerations}
Below, we initiate some theoretical considerations to shed light on the seemingly distinct assembly behaviors of \ch{COOH}-PEG-AuNPs and \ch{NH2}-PEG-AuNPs in response to varying pH and ionic strength.  We use the Poisson-Boltzman (PB) theory to calculate the actual charge for each PEG-AuNP in accordance to measured $\zeta$ potentials and discuss other competing interactions that are not included in the PB theory.

We assume a gold nanoparticle of core diameter $D_{\rm NP}$ functionalized with PEG with either terminal amine or carboxylic groups. As shown in Table~\ref{tbl:dls-zeta}, the bare, unfunctionalized AuNP is negatively charged. This is expected, since as reported in Ref.~[\!\!\citenum{Doane2010}], it is not entirely possible to make the AuNP electrically neutral. Within PB, the $\zeta$ potential,  normalized by the Boltzmann constant $k_{\rm B}$ and temperature $T$,
is given by:
\begin{equation}\label{Eq:zeta_potential}
\frac{e\zeta}{k_{\rm B} T} =2\frac{\mbox{ sign}(\sigma)}{\hat{\lambda}_{\rm GC}}\frac{\hat{R}}{1+\hat{R}}
\end{equation}
where $\hat{R}=R_{\rm h}/\lambda_{\rm DB}$, $\hat{\lambda}_{\rm GC}=\lambda_{\rm GC}/\lambda_{\rm DB}$ with $R_{\rm h}$ the nanoparticle (hydrodynamic) radius and $\lambda_{\rm GC}$, $\lambda_{\rm DB}$ the Guoy-Chapman and Debye lengths, respectively and $\sigma$ the overall charge density (see Supporting Information for further details). The sign function returns 1, 0, and -1 if the argument $\sigma$ is positive, zero, or negative, respectively.

\begin{figure}[!hbt]
\centering 
\includegraphics[width=\linewidth]{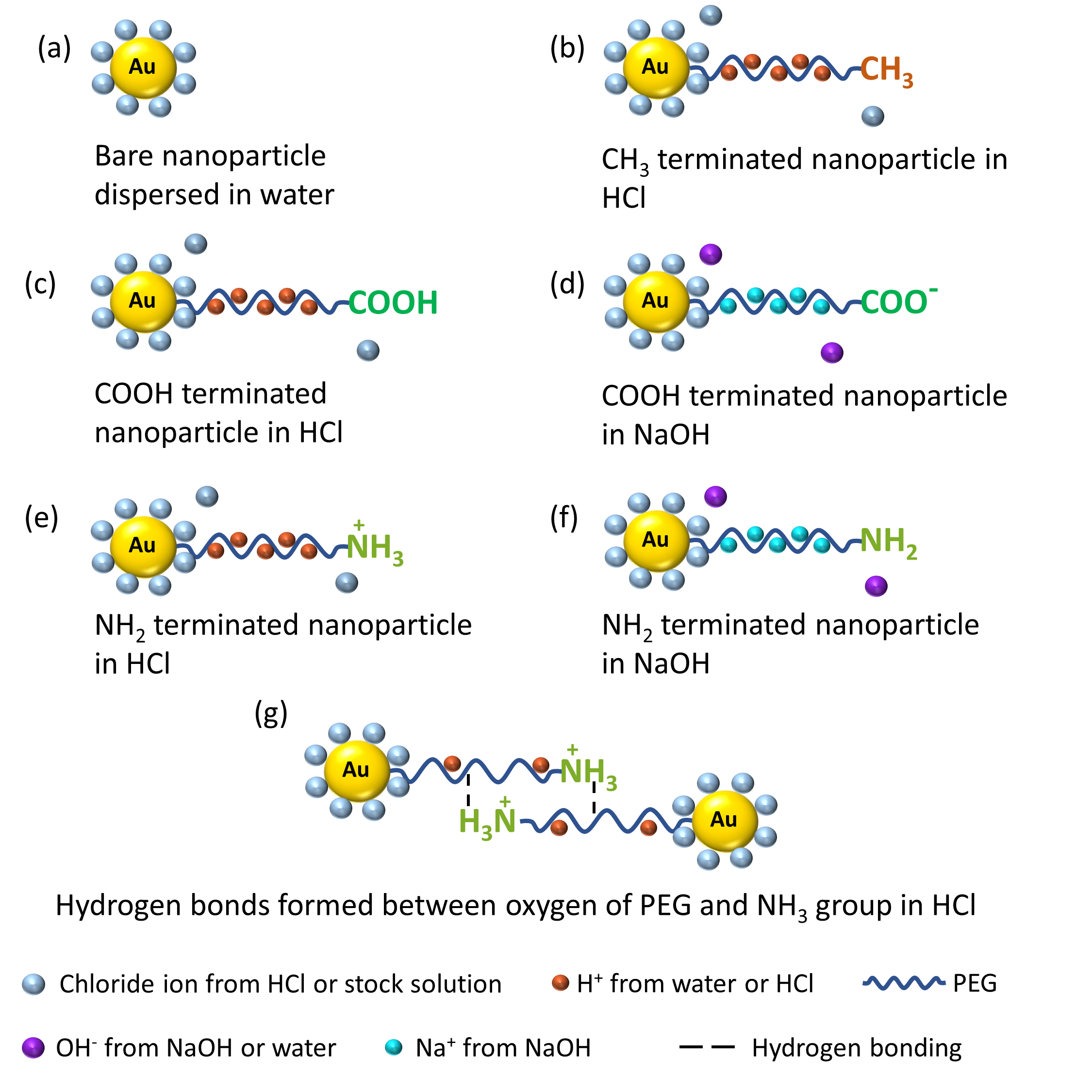}
\caption{ (a) Bare surface AuNPs carry net negative charges due to residual \ch{Cl-} ions.
(b) PEG binds protons, hydroniums and cations, thus appears positively charged and offsets the negative charge of the core AuNP surface.
(c) Both PEG and \ch{COOH} bind proton and appear positively charged in a \ch{H+} abundant environment. However, the carboxyl group is protonated and remain charge neutral. (d) In a high pH environment, PEG binds \ch{Na+} and appears positively charged, but the carboxyl group is deprotonated and carry negative charge, which diminishes the reduction of negative charge of the core AuNP surface by \ch{COOH}-PEG ligands.  
(e) In a low pH environment where \ch{H+} is abundant, both PEG and \ch{NH2} are positively charged, and together greatly offset the negative charge of the AuNPs. (f) In a high pH environment, where \ch{Na+} is abundant, the PEG is positively charged due to \ch{Na+} binding, and \ch{NH2} remains charge neutral. (g) At low pH, there is also highly likely that hydrogen bonds form between \ch{NH2} and PEG, which promotes assembly. 
}
\label{fig:il_1}
\end{figure}

Using the measured values for the nanoparticle hydrodynamic radius (or diameter) and the $\zeta$ potential, we calculate the net charge (from the Guoy-Chapman length) per NP for each case. We then use this information to draw some qualitative considerations.  The charge state of the PEG-AuNPs are conceptually shown in Fig.\ref{fig:il_1}. The total net charge per AuNP obtained based on Table~\ref{tbl:dls-zeta} gives respectively:$-45${\it e} per bare AuNP, $-17${\it e} per \ch{CH3}-PEG-AuNP, $-27${\it e} per \ch{COOH}-PEG-AuNP, and $+16${\it e} per \ch{NH2}-PEG-AuNP, where $e$ denotes the elementary charge. 

The net negative charge present in the gold core, as demonstrated in Fig. \ref{fig:il_1} (a) is likely due to chlorine ions that are left over from the AuNPs growth process \cite{Doane2010,doane2012nanoparticle} and is also inferred in an earlier X-ray reflectivity study \cite{wang2016assembling}. After grafting with PEG, the negative charge of the core is offset by approximately 28 protons that are hydrogen-bonded to the ether oxygen of PEG. The apparent positive charge of PEG, demonstrated in Fig. \ref{fig:il_1} (b), supports the idea that PEG acts like a polyelectrolyte by sequestration of cations.\cite{cao2022poly,fang2017complexing,hakem2004binding} The terminal carboxylic group adds about  $10${\it e} to that of neutral PEG, at near neutral pH, as it is protonated, as is illustrated in Fig. \ref{fig:il_1} (c). Finally, for the PEG terminated with amine there are approximately $+33${\it e} added compared to the PEG with methyl terminal, as is illustrated in Fig. \ref{fig:il_1} (e). 

The PEG chains and their \ch{COOH} terminals act as \ch{H+} acceptors and donors at near neutral pH and thus bind to each other,  consistent with the view that the \ch{COOH}-PEG corona partially collapses rendering the grafted particle a smaller effective hydrodynamic diameter compared to those of the \ch{CH3}- and \ch{NH2}-PEG capped PEG-AuNPs.  At low pH, \ch{COOH} stays charge-neutral while PEG becomes positively charged by acquiring protons or hydronium ions, thus offsets the negative charge of the AuNPs, and assembly of particles is favored.  At high pH, i.e., high [\ch{NaOH}], \ch{COOH} is deprotonated and highly charged, and this additional negative charge offsets the PEG's positive charge associated with bound \ch{Na+}, thus the dispersion state of particles is favored. 
 
On the other hand, \ch{NH2} terminal, in and of itself, is a proton acceptor that enhances total number of positive protons to offset the negative surface charge of the AuNPs.  At low pH where positive protons and hydronium ions are abundant, the positively charged \ch{NH2} terminals, along with positive PEG, greatly offset the intrinsic residual negative charge of AuNPs (see Fig. \ref{fig:il_1} (d)).   At high pH where \ch{NH2} terminal remain charge-neutral, PEG still harvest \ch{Na+} in an \ch{Na+} rich environment and thus carry positive charge (see Fig. \ref{fig:il_1} (e)), thus still offset the negative charge of the AuNP significantly. Therefore, both PEG and \ch{NH2} can together reduce the negative charge of AuNPs, and even inverse the charge polarity from negative to positive, which explains why \ch{NH2}-PEG-AuNPs favors assembly in a much wider pH range than \ch{COOH}-PEG-AuNPs.  

\begin{figure}[!ht]
\centering 
\includegraphics[width=\linewidth]{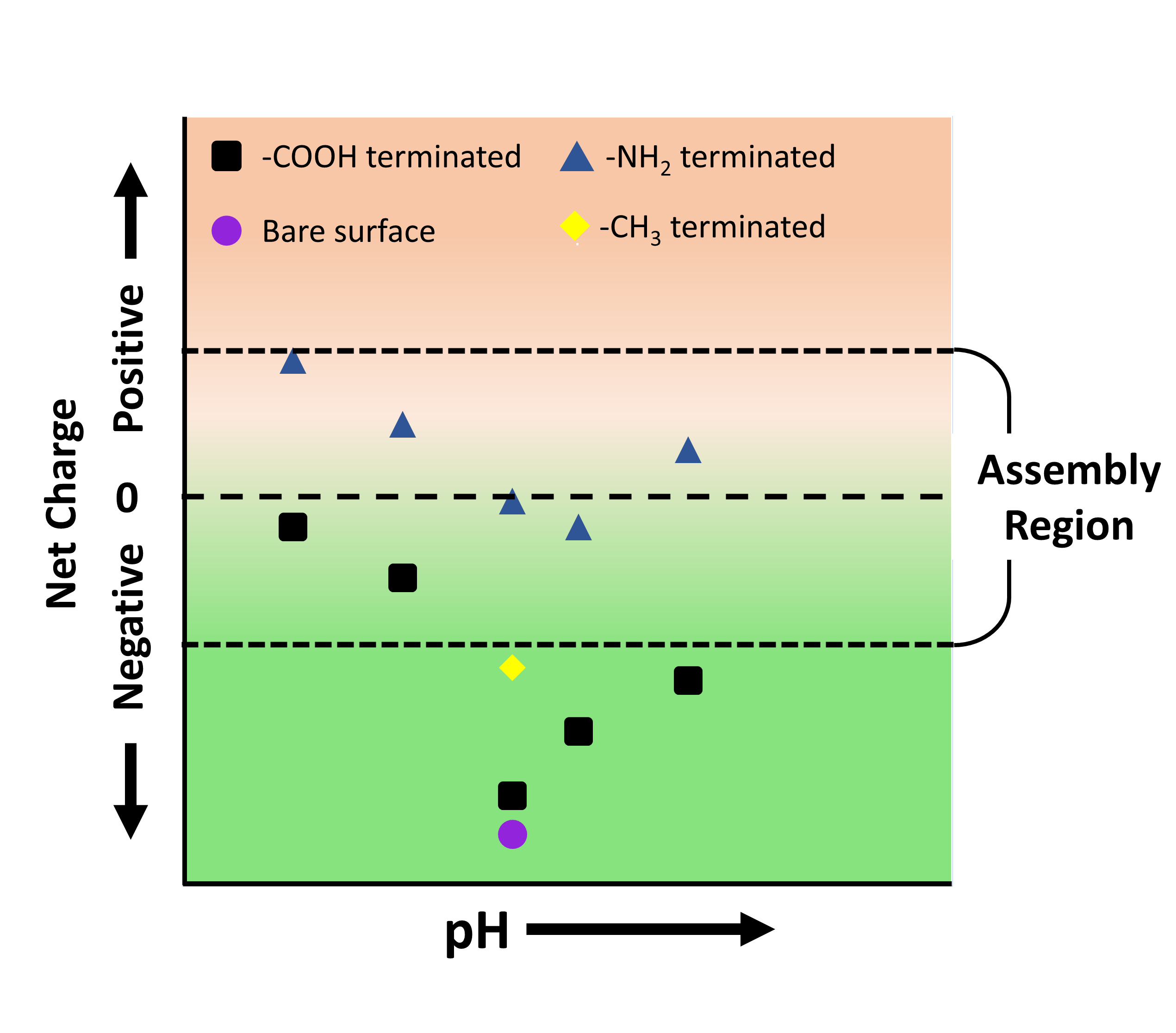}
\caption{ Conceptual illustration for the easiness of assembly of PEG-AuNPs regulated by pH. When the net charge of the PEG-AuNP near zero (i.e., charge neutral), they assemble, and disperse otherwise. 
}
\label{fig:il_2}
\end{figure}
 
Thus, our theoretical considerations to rationalize the results in the framework of the Poisson-Boltzmann theory explain the assembly behavior of both the \ch{COOH}- and \ch{NH2}-PEG capped AuNPs. It is worth noting that \ch{NH2} terminals also form hydrogen bonds with the ether oxygen in PEG at low pH when proton is abundant, as shown in Fig. \ref{fig:il_1} (g). This may be the main reason why charged \ch{NH2} terminals also promote assembly regardless of pH.  A more thorough molecular simulation study is underway to shed more light on this regard.     

The PB based perspective also helps understand the assembly driven by \ch{K2CO3}, when cations concentration, [\ch{K+}], is orders of magnitude higher than that of \ch{Na+} at  comparable pH levels (see Fig. \ref{fig:gamma_COOH}). Even though \ch{COOH} terminals are deprotonated and thus negatively charged, PEG may acquire enough \ch{K+} to neutralize the whole particle to trigger assembly.
Figure \ref{fig:il_2} provides a conceptual illustration of PB-theory based understanding. The \ch{NH2}-PEG harvest positive charge that greatly offset the intrinsic negative charge of the AuNPs when there is abundance of \ch{H^+} (i.e., low pH) and \ch{Na^{+}} (i.e., high pH) in the solution, resulting with a neutral NP prone to assembly.  While COOH-PEG only offset AuNP negative surface charge with abundance of \ch{H^+} to assemble. At high pH, the deprotonated carboxylic group carry negative charge, partially cancels out the PEG positive charge arising from the bound \ch{Na^+} ions. 



\section{Conclusions}
NPs decorated with opposite surface charges can constitute  building blocks of ionic colloidal crystals yielding another control knob to achieve super-stoichiometry.  As we embark on assembling NPs of opposite charges, we start by studying the properties and assembly of PEG-AuNPs of like charge. Here, we report the results of ligand PEG with various terminal groups, i.e., \ch{CH3} (neutral), \ch{NH2} (positively charged), and \ch{COOH} (negatively charged).  Dynamic light scattering measurements show that the  hydrodynamic diameter of \ch{NH2}-PEG-AuNP (core and PEG-corona) is slightly larger ($\sim 5\%$) than that  of \ch{CH3}-PEG-AuNP,  while \ch{COOH}-PEG-AuNP is significantly smaller ($\sim 18\%$). The $\zeta$ potentials are associated with the expected charge for the terminal groups.  The net charge of the grafted AuNPs affect their assembly at the vapor/liquid interface. Using surface sensitive synchrotron X-ray diffraction techniques, we find that the grafted AuNPs with \ch{NH2} spontaneously migrate to the surface and form a two-dimensional hexagonal structure with similar lattice constant to one obtained  with a neutral terminal group \ch{CH3} in the presence \ch{K2CO3}.  The \ch{COOH}-PEG-AuNPs migrate and form 2D hexagonal structure on the aqueous surface only  at moderately high salt (100 mM \ch{K2CO3}) concentrations, with significantly smaller lattice constant. The aqueous pH regulates the assembly and dispersion behavior of \ch{COOH}-PEG-AuNPs but has nearly no effect on that of \ch{NH2}-PEG-AuNPs.  
The difference in the assembly behavior of \ch{COOH}-PEG-AuNPs and \ch{NH2}-PEG-AuNPs in response to pH and salinity regulation is attributed to the intrinsic residual negative charge on AuNP's surface and PEG's sequestration of cations that make up a significant portion of the net charge, other than the charge of the ligand terminals upon protonation or deprotonation.  The fact that \ch{NH2}-PEG-AuNPs readily assemble even at low pH suggests that the coupling interaction between \ch{NH2} terminal group and PEG, most likely hydrogen bonding, may promote spontaneous assembly.  Our results, dealing with charged AuNPs, pave the way to ongoing future work to assemble oppositely charged grafted NPs into binary super-ionic crystals analogous to atomistic ionic crystals. 



\section{Acknowledgements}
Research was supported by the U.S. Department of Energy (U.S. DOE), Office of Basic Energy Sciences, Division of Materials Sciences and Engineering. Ames National Laboratory is operated for the U.S. DOE by Iowa State University under Contract DE-AC02-07CH11358. 
This research used the Open Platform Liquid Surfaces (OPLS) end station of the Soft Matter Interfaces Beamline (SMI, Beamline 12-ID)  at the National Synchrotron Light Source II, a U.S. Department of Energy (DOE) Office of Science User Facility operated for the DOE Office of Science by Brookhaven National Laboratory under Contract No. DE-SC0012704.

\normalem
\bibliography{Reference.bib}
\clearpage
\onecolumn
 
\clearpage
\onecolumn

\setcounter{page}{1}
\setcounter{figure}{0}
\setcounter{equation}{0}
\setcounter{table}{0}

\renewcommand{\thefigure}{S\arabic{figure}}
\renewcommand{\theequation}{S\arabic{equation}}
\renewcommand{\thetable}{S\arabic{table}}
\renewcommand{\thepage}{S\arabic{page}} 

\section{{\Large Supporting information}}
\LARGE

\begin{center} 
	{\bf Two-dimensional assembly of nanoparticles grafted with charged-end-group polymers}\\
	\bigskip
	\normalsize
	Hyeong Jin Kim,$^\dagger$ 
	Binay P. Nayak,$^\dagger$
	Honghu Zhang,$^\ddagger$
	Benjamin M. Ocko,$^\mathparagraph$ 
	David Vaknin,$^{\S}$
	Alex Travesset,$^{\S}$
	Surya K. Mallapragada,$^{\ast ,\dagger}$ and Wenjie Wang,$^{\ast ,\mid\mid}$\\
	\bigskip
	{$\dagger$\it Ames National Laboratory, and Department of Chemical and Biological Engineering, Iowa State University, Ames, Iowa 50011, United States}\\
	{$\ddagger$\it NSLS-II, Brookhaven National Laboratory, Upton, New York 11973, United States }\\
	{$\mathparagraph$\it Center for Functional Nanomaterials and NSLS-II, Brookhaven National Laboratory, Upton, New York 11973, United States}\\
	{$\S$\it Ames National Laboratory, and Department of Physics and Astronomy, Iowa State University, Ames, Iowa 50011, United States}\\
	{$\mid\mid$\it Division of Materials Sciences and Engineering, Ames National Laboratory, U.S. DOE, Ames, Iowa 50011, United States}\\
	\bigskip
	{E-mail: suryakm@iastate.edu; wenjiew@ameslab.gov}\\
\end{center}

\subsection{Simple calculation of the pH for \ch{K2CO3} solutions}\normalsize
The \ch{K2CO3} stock solution was prepared
with nominal amount of \ch{K2CO3} dissolved in pure water and further dilute to the nanoparticle suspensions to specified concentration. The pH is determined in terms of known equilibrium constants for room temperature via textbook style calculation, rather than actual measurements.
The \ch{CO3^{2-}} is the conjugate base of the weak acid \ch{HCO3^{-}} with its known acid dissociation constant, $K_a=4.7\times 10^{-11}$, for $K_a$=[\ch{CO3^{2-}}][\ch{H^+}]/[\ch{HCO3^{-}}]. 
So for the equilibrium \ch{CO3^{2-} + H2O <=> HCO3- + OH-}, the corresponding  base dissociation constant $K_b=K_w/K_a=2.12\times10^{-4}$, where $K_w$ is the dissociation constant of water (=$10^{-14}$). Thus, we determine [\ch{OH-}] based on the known [\ch{K2CO3}], assuming [\ch{OH-}]$\approx$[\ch{HCO3-}] and [\ch{HCO3-}]$\ll$[\ch{CO3^{2-}}] in the aqueous solution. 
So the pH equals to $14+\log_{10}(\sqrt{K_b\times[\rm K_2CO_3]})$.
For [\ch{K2CO3}]=1, 10, 100 mM, the corresponding pH=10.7, 11.2, 11,7, respectively.

\subsection{Grafting density from TGA}

\begin{figure} 
    \centering
    \includegraphics[width=0.8\linewidth] {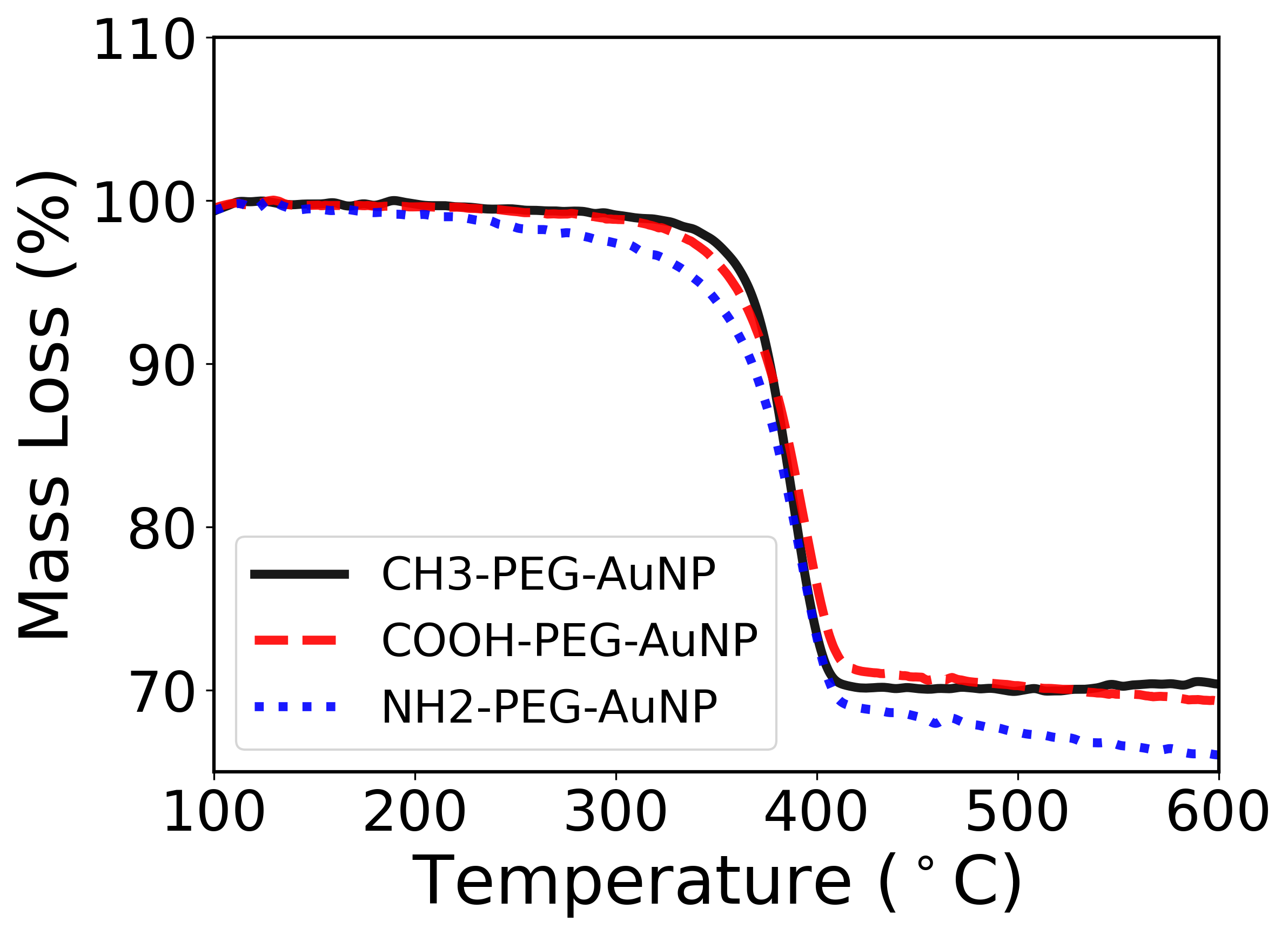}
    \caption{\small  TGA showing relative mass loss percentage as a function of the temperature of polymer grafted AuNPs.}
\label{fig:TGA}
\end{figure}

To calculate the grafting density from TGA using Eq.~\ref{Eq:TGA}, first, the relative mass of the PEG shell and gold nanoparticles was calculated from Fig.~\ref{fig:TGA}. The total number of polymer chains was calculated by dividing the weight of the PEG from the experiment by the mass of each chain of PEG which is 5 kDa or $8.2027\times10^{-21}$ gm.
The total number of nanoparticles was calculated by dividing the weight of the nanoparticles from the experiment with the mass of individual nanoparticles calculated by multiplying the volume of each nanoparticle, assuming it is to be in the spherical shape of diameter 9.34 nm (calculated from form factor) and having a density of 19.6 g/cm$^3$. Since we are calculating the grafting density as chains per nm$^2$, we need to divide the total area of nanoparticles, which is calculated again, assuming it to be the spherical shape of diameter 9.34 nm.

\begin{equation}\label{Eq:TGA}
   \sigma_{\rm TGA} = \frac{{\rho}_{\rm NP}}{6}\times \frac{\rm PEG\; wt\%}{\rm NP\; wt\%}
    \times \frac{D_{\rm NP}}{\mathrm{MW}_{\rm PEG}}.
\end{equation}
where $\sigma_{\rm TGA}$ is the grafting density,
$\mathrm{wt}\% $ is the calculated change in relative weight,
$\rho_{\rm NP}$ is the mass density of gold, and $\mathrm{MW}_{\rm PEG} $ represents the molecular weight of PEG.
\subsection{Electrostatic Calculation: Poisson Boltzmann equation}

Using the Poisson-Boltzmann formalism and the known pKa values for the terminal groups we calculate surface charges and $\zeta$-potential for the two charged terminal groups. 
The electric potential $\psi(r)$ is related to electric field as follows, 
\begin{equation}\label{Eq:PB:electric_field}
    E(r)=-\frac{d\psi(r)}{dr} \ .
\end{equation}

There are $a=1 \cdots f$ ion species (with charge $e q_a$), with number density given by Boltzmann statistics 
\begin{equation}\label{Eq:PB:density}
    n_a(r)=\frac{1}{\upsilon} e^{\frac{\mu_a}{k_B T}} e^{-q_a \frac{e \psi(r)}{k_B T}} \ ,
\end{equation}
where $\upsilon$ is a reference volume and $\mu_a$ the corresponding chemical potential. The Poisson-Boltzmann equation is

\begin{eqnarray}
\label{Eq:PB:equation}
    \frac{1}{r^2}\frac{d}{dr}\left(r^2 \frac{d \psi(r)}{dr} \right)= -\frac{e}{\varepsilon_0 \varepsilon_w}\sum_{a=1}^f q_a n_a(r) 
 \nonumber \\ =-\frac{e}{\varepsilon_0 \varepsilon_w}\sum_{a=1}^f q_a n_a^{\infty} e^{-q_a e \frac{\psi(r)}{k_B T}} .
\end{eqnarray}

The above equation is
\begin{equation}\label{Eq:PB:equation_electric}
     \frac{1}{r^2}\frac{d}{dr}\left(r^2 E(r) \right)=\frac{e}{\varepsilon_0 \varepsilon_w}\sum_{a=1}^f q_a n_a^{\infty} e^{-q_a e \frac{\psi(r)}{k_B T}} .
\end{equation}
which implies, with $R$ the NP radius with a wall located at a distance $L$ 
\begin{eqnarray}
\label{Eq:PB:equation_electric_explicit}
     L^2 E(L)-r^2 E(r) \nonumber   \\ =\frac{e}{\varepsilon_0 \varepsilon_w}\int_r^{L} r^2 dr\sum_{a=1}^f q_a n_a^{\infty} e^{-q_a e \frac{\psi(r)}{k_B T}} .
\end{eqnarray}
It is $E(L)=0$, hence
\begin{eqnarray}
\label{Eq:PB:bc}
    -R^2 E(R)=\frac{e}{\varepsilon_0 \varepsilon_w} \int_{L}^{R} r^2 dr\sum_{a=1}^f q_a n_a^{\infty} e^{-q_a e \frac{\psi(r)}{k_B T}} \nonumber \\ \nonumber=-\frac{\sigma}{\varepsilon_0 \varepsilon_w} R^2 \rightarrow E(R)=\frac{\sigma}{\varepsilon_0 \varepsilon_w} \ .
\end{eqnarray}

In dimensionless variables $u=r/\lambda_{DB}$, $\phi=\frac{e \psi}{k_B T}$
where $\lambda_{DB}^2=1/(8\pi l_B I)$, with $l_B$ the Bjerrum length and $I$ the ionic strength. The PB equation is
\begin{equation}\label{Eq:PB:PB_equation_dimensionless}
    \frac{1}{u^2}\frac{d}{du}\left(u^2 \frac{d\phi(u)}{du}\right)=-\sum_{a=1} \frac{q_a n_a^{\infty}}{2 I}e^{-q_a\phi(u)} \ , 
\end{equation}
with boundary condition
\begin{equation}\label{Eq:PB:bc_dimensionless}
    \frac{d\phi}{du}=-2\mbox{ sign}(\sigma) \frac{\lambda_{DB}}{\lambda_{GC}} \ .
\end{equation}
where $\lambda_{GC}=\frac{e}{2\pi l_B \sigma}$ is the Guoy-Champan length. 
Within the linear approximation
\begin{equation}\label{Eq:PB:equation_linear}
    \frac{1}{u^2}\frac{d}{du}\left(u^2 \frac{d \phi(u)}{du} \right)= \phi(u) \ .
\end{equation}
The solution for the potential and electric field is
\begin{eqnarray}\label{Eq:PB:equation_linear_solution}
    \phi(u)&=&2\mbox{ sign}(\sigma)\frac{\hat{R}^2}{\hat{\lambda}_{GC}}\frac{1-\frac{\hat{L}+1}{\hat{L}-1}e^{-2(\hat{L}-u)}}{1-\frac{(\hat{R}-1)(\hat{L}+1)}{(\hat{R}+1)(\hat{L}-1)}e^{-2(\hat{L}-\hat{R})}}\frac{e^{-(u-\hat{R})}}{u(1+\hat{R})} \\\nonumber
    E(u)&=& 2\mbox{ sign}(\sigma)\frac{\hat{R}^2}{\hat{\lambda}_{GC}}\frac{u+1-(u-1)\frac{\hat{L}+1}{\hat{L}-1}e^{-2(\hat{L}-u)}}{1-\frac{(\hat{R}-1)(\hat{L}+1)}{(\hat{R}+1)(\hat{L}-1)}e^{-2(\hat{L}-\hat{R})}}\frac{e^{-(u-\hat{R})}}{u^2(1+\hat{R})}
\end{eqnarray}
$\hat{R}=\frac{R}{\lambda_{DB}}$, $\hat{L}=\frac{L}{\lambda_{DB}}$ and $\hat{\lambda}_{GC}=\frac{\lambda_{GC}}{\lambda_{DB}}$.

Comparing the linear solution to the exact one from Eq.~\ref{Eq:PB:PB_equation_dimensionless} the difference for the situations considered here is within 5\% at worst (at zero salt concentration, pH=7). Hence, there is negligible error in using the linear solution and keeping the ion distribution as in Eq.~\ref{Eq:PB:density}, that is, without Taylor-expanding the exponential.

\end{document}